\documentclass[12pt]{article}
\usepackage{amsmath,amsfonts,amssymb,a4,epsfig}

\newcommand{\R}{{\mathbb{R}}}

\newcommand{\Z}{{\mathbb{Z}}}
\newcommand{\N}{{\mathbb{N}}}

\def\ha{\frac{1}{2}}

\def\pa{\partial}
\def\ra{\rightarrow}
\def\preuve{\begin{proof}} 
\def\ga{\alpha}
\def\gb{\beta}

\def\gl{\lambda}
\def\go{\omega}

\def\san{San V{\~u} Ng{\d o}c}

\newtheorem{defi}{Definition}[section]

\newtheorem{theo}{Theorem}[section]

\def\YCV{Yves  Colin de Verdi{\`e}re}

\begin{document}

\title{The semi-classical
spectrum \\  and  \\
the Birkhoff normal form}
\author{Yves Colin de Verdi\`ere\footnote{Institut Fourier,
 Unit{\'e} mixte
 de recherche CNRS-UJF 5582,
 BP 74, 38402-Saint Martin d'H\`eres Cedex (France);
yves.colin-de-verdiere@ujf-grenoble.fr}}
\maketitle
\section*{Introduction}

The purposes of this note are
\begin{itemize}
\item To propose  a direct and ``elementary'' proof of the main result of
  \cite{GPU}, namely that the semi-classical spectrum near a global
minimum of the classical Hamiltonian determines the whole
semi-classical
Birkhoff normal form (denoted the  BNF) in the non-resonant case.
I believe  however that the method used in \cite{GPU} (trace formulas)
are more general and can be applied to any non degenerate
non resonant critical point  provided that the corresponding
critical value is ``simple''.

\item To present in the completely resonant case a similar problem
which is NOT what is done in \cite{GPU}: there, only the {\it non-resonant
part} of the BNF is proved to be determined from the semi-classical spectrum!
\end{itemize}
\section{A direct proof of the main result of \cite{GPU}}
\subsection{The Theorem}
Let us give a semi-classical  Hamiltonian $\hat{H}$
  on $\R^{d}$ (or even on a smooth connected
manifold of dimension $d$) which is the Weyl quantization 
of the symbol  $H\equiv H_0 +\hbar H_1 + \hbar^2 H_2 +\cdots $.

Let us assume that $H_0 $ has a global non degenerate non
resonant minimum $E_0$  at the point $z_0$:
it means that after some affine  symplectic change
of variables
$H_0=E_0+ \ha  \sum _{j=1}^d \go_j (x_j^2+\xi_j^2)+ \cdots $
where the $\go_j $'s are $>0$ and  independent over the rationals.
We can assume that $0<\go_1 <  \go _2 <  \cdots < \go_d$.
We will denote $E_1=H_1(z_0)$. 

We assume also that
\[ \liminf _{(x,\xi)\ra \infty}H (x,\xi)> E_0 ~.\]

Let us denote by $\gl_1 (\hbar)< \gl_2 (\hbar ) \leq \cdots 
\leq \gl_N (\hbar ) \leq \cdots $
the discrete spectrum of $\hat{H}$. This set can be finite for a 
fixed value of $\hbar $, but, if $N$ is given,
$ \gl_N(\hbar)$ exists for $\hbar $ small enough. 
\begin{defi}
The {\rm semi-classical spectrum}  of $\hat{H}$ is the set
of all $\gl_N(\hbar)$ ($N=1,\cdots $) modulo $O(\hbar^\infty )$.
{\bf NO uniformity with respect to $N$ in the $O(\hbar^\infty )$ 
is required.}
\end{defi}

\begin{defi}
The semi-classical Birkhoff normal form is
the following formal series expansion in $\Omega
=(\Omega _1, \cdots , \Omega _d )$ and $\hbar$:
\[ \hat{B}\equiv E_0 + \hbar E_1 +  \sum_{j=1}^d  \go_j \Omega _j
+ \sum_{l+|\ga| \geq 2 } c_{l,\ga }\hbar^l \Omega ^\ga \]
with  $\Omega _j=\ha \left( -\hbar^2 \pa_j ^2 +x_j^2 \right) $.
The series 
$\hat{B}$ is uniquely defined as being the Weyl quantization
of  some symbol $B$ equivalent to the Taylor
expansion
at $z_0$ of  ${H}$
by some automorphism of the semi-classical Weyl algebra (see \cite{YCdV}).
\end{defi}

The main result is the 
\begin{theo}[\cite{GPU}]\label{Main}
Assume as before that the $\go_j$'s are linearly independent
 over the rationals.
Then the semi-classical spectrum
and the semi-classical Birkhoff normal form
determine each other.
\end{theo}

The main difficulty is that the spectrum of $\hat{B}$
is naturally labelled by $d-$uples ${\bf k}\in\Z_+^d$
while the semi-classical spectrum is labelled
by $N\in \N $.
We will denote by $\psi $ the bijection 
$\psi : N \ra {\bf k} $ of $\N $ onto
$\Z_+^d:=\{ {\bf k}=(k_1,\cdots ,k_d)|\forall j,~k_j \in \Z,~k_j \geq 0 \}$
 given by ordering  the numbers
$\langle \go| {\bf k} \rangle $ in increasing order:
they are pair-wise distincts because of the non-resonant
assumption.

\section{From the semi-classical Birkhoff normal form
to the semi-classical spectrum}

We have the following result
\begin{theo}
The semi-classical spectrum is given
by the following power series in $\hbar $:
\begin{equation} \label{equ:main}  \gl_N(\hbar)\equiv
E_0+\hbar \left( E_1 + \ha \langle \go |\psi(N) +\ha \rangle \right) 
+\sum _{j=2}^\infty \hbar^j P_j (\psi(N)) \end{equation}
where the $P_j$'s are polynomials of degree $j$
given by
\[ P_j(Z)=\sum_{l+|\ga |=j}
c_{l,\ga }\left(Z+\ha \right)^\ga ~.\]
\end{theo}

This result is an immediate consequence
of results proved  by B. Simon \cite{Si}  and 
B. Helffer-J. Sj\"ostrand \cite{H-S} concerning
the first terms,  and by J. Sj\"ostrand in \cite{Sj} (Theorem 0.1)
where he proved a much stronger result.

\section{From the semi-classical spectrum to the $\omega _j$'s}
\subsection{Determining the $\go_j$'s}\label{sec:omega_j}
Because $E_0=\lim _{\hbar\ra 0}\gl_1 (\hbar) $, we can substract $E_0$
and assume $E_0=0$.

By looking at the limits, 
as $\hbar \ra 0$, $\mu_N:= \lim \gl_N(\hbar)/\hbar$ ($N$ fixed),
 we know the set of all
$E_1 + \sum _{j=1}^d \go_j (k_j +\ha),~(k_1,\cdots, k_d)\in\Z_+^d$.

{\bf Let us give 2 proofs that the $\mu_N$'s determine
the $\go_j$'s.}

\begin{enumerate} 
\item {\bf Using the partition function:}
from the $\mu_N$'s, we know  the meromorphic function
\[ Z(z):=\sum e^{-z \mu_N } ~.\] 
\[ Z(z):=e^{-z(E_1+\ha \sum_{j=1}^d  \go_j )}\sum_{{\bf k}\in \Z_+^d }
 e^{-z\langle {\bf \go}|{\bf k  } \rangle }~, \]

We have
\[ Z(z)=e^{-z(E_1+ \ha \sum_{j=1}^d \go _j )}
\Pi _{j=1}^d (1-e^{-z\go_j })^{-1} ~,\]

The poles of $Z$ are 
${\cal P}:=\cup _{j=1,\cdots, d} \{ \frac{2\pi i \Z}{\go_j }\} $.
The set of $\go_j $ is hence determined up to 
a permutation.
We fix now ${\bf \go}=(\go_1,\cdots, \go_d)$
with $\go_1< \go_2 <\cdots $.

From the knowledge of the $\go_j$'s, we get the bijection $\psi$.
\item {\bf A more elementary proof:}
substract $\mu_1 =E_1+\ha \sum \go_j $ from the whole sequence
and denote $\nu_N=\mu_N -\mu_1 $.
Then $\go_1=\nu_2$. Then remove the multiples of $\go_1$.
The first remaining term is $\go_2$. Remove
all integer linear combinations of $\go_1$ and $\go_2$, the first
remaining term is $\go_3$, $\cdots $
\end{enumerate}

\subsection{Determining the $c_{l,\ga}$'s}

Let us first fix $N$: 
from  Equation (\ref{equ:main}) and 
the knowledge of $\gl_N $ mod $O(\hbar^\infty )$ we know
the $P_j(\psi (N))$'s for all $j$'s.

Doing that for all $N$'s  and using $\psi $
determine the restriction of the $P_j$'s to $\Z_+^d$
and hence the $P_j$'s.

\section{A natural question in the resonant case}

\subsection{The context}

For simplicity, we will consider the completely resonant
case $\go_1=\cdots = \go_d=1$ and work with the Weyl symbols.
 Let us denote by
${\Sigma }=\ha \sum (x_j^2+\xi_j^2) $.

The (Weyl symbol of the)  QBNF is then of the form 
\[ {B}\equiv {\Sigma } + \hbar P_{0,1}
   +\sum_{n=2}^\infty  \sum_{j+l =n} \hbar ^{j} {P}_{2l,j} \] 
where  $P_{2l,j}$ is an
homogeneous polynomial of degree $2l$ in $(x,\xi)$,
Poisson commuting with $\Sigma $:
 $\{ {\Sigma },  P_{2l,j}\}=0 $\footnote{
The Moyal bracket of any  $A$ with $H_2$ reduces to the Poisson
bracket}.

For example,
the first non trivial terms are:
\begin{itemize}
\item for $n=2$: $ P_{4,0} +\hbar P_{2,1}+ \hbar^2 P_{0,2} $
\item   for $n=3$: $P_{6,0}+\hbar P_{4,1}  +\hbar^2 P_{2,2}
+ \hbar^3 P_{0,3} $.
\end{itemize}

The semi-classical spectrum splits into clusters $C_N$ of 
 $N+1$ eigenvalues in an interval of 
size $O(\hbar^2)$  around each $\hbar (N+\ha d +P_{0,1} )$ with $N=0,1,\cdots $.

The whole series $B$ is however NOT unique,
contrary to the non-resonant case, 
but defined up to automorphism of the semi-classical
Weyl algebra  commuting with $\Sigma  $.

Let  $G$ be the group of such automorphisms  (see \cite{YCdV}).
 The natural question is roughly:

{\bf Is the QBNF determined modulo $G$ from the semi-classical
spectrum, i.e. from all the clusters?}

\subsection{The group $G$}

The linear part of $G$ is the group $M$ of all  $A$'s in the  symplectic group
which commute with $\hat{H}_2 $, i.e. the unitary group
$U(d)$.

We have an exact sequence of groups:
\[ 0 \ra K \ra G \ra M \ra 0 ~.\]

Let us describe $K$ (the ``pseudo-differential'' part):

Let $S=S_3 + \cdots $ in  the Weyl algebra
(the formal power series in $(\hbar,x,\xi)$ with the Moyal 
product and the usual grading
${\rm degree}(\hbar^{j} x^\ga \xi^\gb) =2j+|\ga|+|\gb| $)
\[ g_S (H)=e^{iS/\hbar }\star H \star  e^{-iS/\hbar } \]
preserves $\Sigma$ iff $\{ S_n, \Sigma \}=0 $.
This implies that $n$ is even and 
$S_{n }$ is a polynomial in
$z_j\overline{z_k }$ ($z_j=x_j+i\xi_j $).
Then $K$ is the group of all $g_S$'s with
$\{ S, \Sigma \}=0$.

\bibliographystyle{plain}

\begin{thebibliography}{99}

\bibitem{San-C} Laurent Charles   \&  \san.
Spectral asymptotics via the Birkhoff normal form.
{\it ArXiv:math-SP/0605096},  {\it Duke Math. Journal} {\bf
    143}:463--511 (2008).

\bibitem{YCdV} \YCV.
An extension of the Duistermaat-Singer Theorem to
the semi-classical Weyl algebra.{\it  Preprint 2008.}

\bibitem{GPU} Victor Guillemin, Thierry  Paul \& Alejandro Uribe.
``Bottom of the well'' semi-classical wave trace invariants.
{\it ArXiv:math-SP/0608617 {\rm and}
Math. Res. Lett. 14:711--719 (2007).}

\bibitem{H-S} Bernard Helffer \& Johannes Sj\"ostrand.
Puis multiples en semi-classique I.
{\it Commun. PDE.} {\bf 9}:337--408 (1984).


\bibitem{Si} Barry Simon. Semi-classical analysis of low lying
  eigenvalues I: Non degenerate minima.
{\it Ann. IHP (phys. th\'eo.)} {\bf 38}: 295--307 (1983).




\bibitem{Sj}
Johannes Sj\"ostrand.
Semi-excited states  in nondegenerate potential wells.
{\it Asymptotic Analysis} {\bf 6}:29--43 (1992).



\end{thebibliography}

\end{document}